\documentclass[preprint,journal]{vgtc}       

\ifpdf
  \pdfoutput=1\relax                   
  \usepackage{graphicx}                
  \DeclareGraphicsExtensions{.pdf,.png,.jpg,.jpeg} 
\else
  \usepackage{graphicx}                
  \DeclareGraphicsExtensions{.eps}     
\fi%

\graphicspath{{figures/}{pictures/}{images/}{./}} 

\usepackage{microtype}                 
\PassOptionsToPackage{warn}{textcomp}  
\usepackage{textcomp}                  
\usepackage{mathptmx}                  
\usepackage{times}                     
\usepackage{cite}                      
\usepackage{tabu}                      
\usepackage{booktabs}                  
\usepackage{xspace}
\usepackage{comment}
\usepackage{enumitem}
\PassOptionsToPackage{hyphens}{url}\usepackage{hyperref}

\onlineid{1122}

\vgtccategory{Research}
\vgtcpapertype{system}

\usepackage[dvipsnames]{xcolor}
\usepackage[normalem]{ulem}
\definecolor{mred}{rgb}{.80,.12,.30}
\definecolor{MRED}{rgb}{.80,.12,.30}
\definecolor{grey}{rgb}{0.5,0.5,0.5}
\definecolor{purple}{rgb}{.75,0,.85}
\definecolor{pistachio}{rgb}{0.58, 0.77, 0.45}

\newif\ifnotes
\notestrue

\let\origcite\cite
\renewcommand{\cite}[1]{\ifnotes\mbox{\origcite{#1}}\else \origcite{#1}\fi}

\newcommand{\sys}{CAVA\xspace}
\newcommand{\userML}{Andy\xspace}

\title{CAVA: A Visual Analytics System for Exploratory Columnar Data Augmentation Using Knowledge Graphs}

\author{Dylan Cashman, Shenyu Xu, Subhajit Das, Florian Heimerl, Cong Liu, Shah Rukh Humayoun, \\ Michael Gleicher, Alex Endert, Remco Chang}
\authorfooter{
\item
Dylan Cashman, Cong Liu, and Remco Chang are with Tufts University. E-mail: \{dcashm01, cong, remco\}@cs.tufts.edu.
\item
Shenyu Xu, Subhajit Das, and Alex Endert are with Georgia Tech. E-mail: \{das, shenyuxu, endert\}@gatech.edu.
\item
Florian Heimerl and Michael Gleicher are with University of Wisconsin, Madison.  E-mail: \{heimerl, gleicher\}@cs.wisc.edu.
\item
Shah Rukh Humayoun is at San Francisco State University.  E-mail: humayoun@sfsu.edu.
}

\shortauthortitle{Cashman \MakeLowercase{\textit{et al.}}: CAVA: Columnar Data Augmentation}

\abstract{
Most visual analytics systems assume that all foraging for data happens before the analytics process; once analysis begins, the set of data attributes considered is fixed. Such separation of data construction from analysis precludes iteration that can enable foraging informed by the needs that arise in-situ during the analysis. The separation of the foraging loop from the data analysis tasks can limit the pace and scope of analysis. In this paper, we present \sys, a system that integrates data curation and data augmentation with the traditional data exploration and analysis tasks, enabling information foraging in-situ during analysis. Identifying attributes to add to the dataset is difficult because it requires human knowledge to determine which available attributes will be helpful for the ensuing analytical tasks. \sys crawls knowledge graphs to provide users with a a broad set of attributes drawn from external data to choose from. Users can then specify complex operations on knowledge graphs to construct additional attributes. \sys shows how visual analytics can help users forage for attributes by letting users visually explore the set of available data, and by serving as an interface for query construction.  It also provides visualizations of the knowledge graph itself to help users understand complex joins such as multi-hop aggregations. We assess the ability of our system to enable users to perform complex data combinations without programming in a user study over two datasets. We then demonstrate the generalizability of \sys through two additional usage scenarios. The results of the evaluation confirm that \sys is effective in helping the user perform data foraging that leads to improved analysis outcomes, and offer evidence in support of integrating data augmentation as a part of the visual analytics pipeline.
} 


\teaser{
 \centering
     \vspace{-2mm}

 \includegraphics[width=\linewidth]{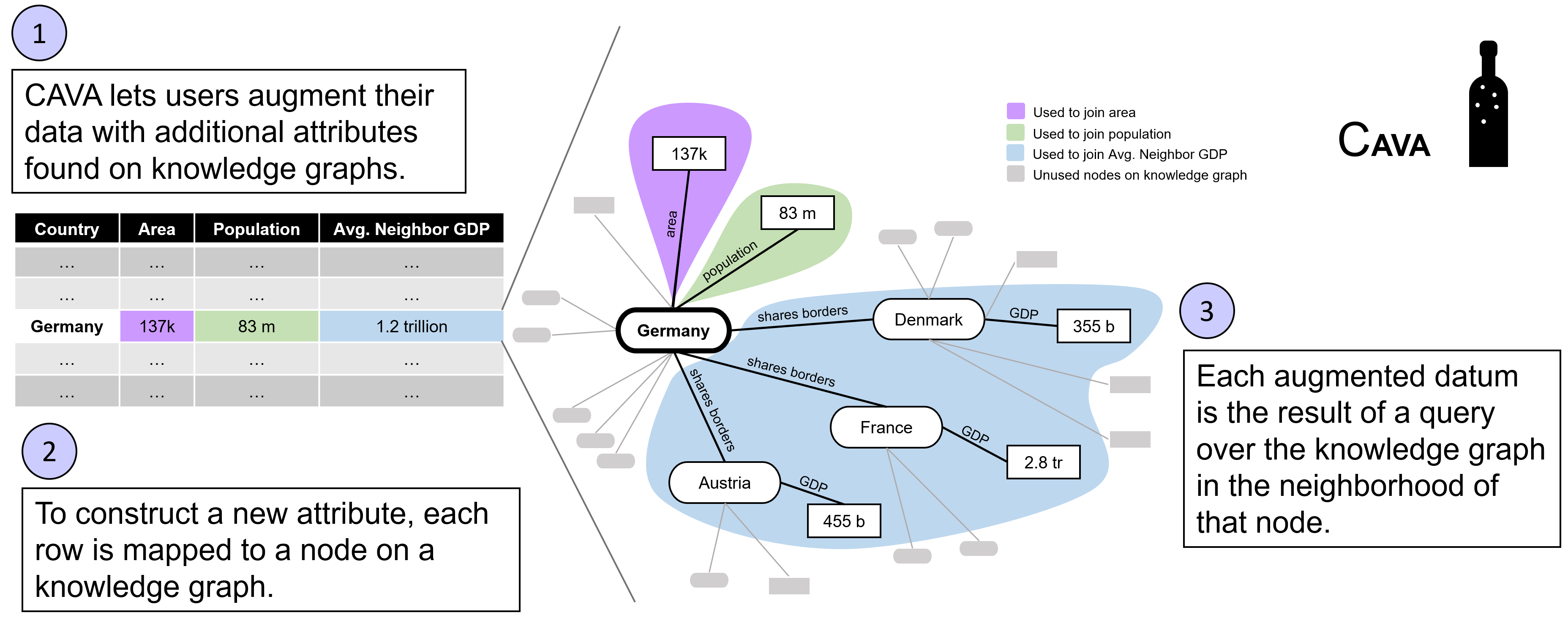}
 \caption{The \sys system allows a user to augment a tabular dataset with additional attributes gathered from a knowledge graph.  This figure illustrates the process of gathering three additional attributes corresponding for a single row of the data. }
     \vspace{-1mm}
	\label{fig:teaser}
}

\vgtcinsertpkg

\begin{document}


\firstsection{Introduction}

\maketitle

Over 20 years ago, Pirolli and Card coined the term ``information foraging'' to denote the process of seeking and gathering information to apply towards a task\cite{pirolli1999information, pirolli2005sensemaking}.  While they focused on foraging for additional entities or rows of a dataset, foraging for additional data attributes or columns can unlock new analytical capabilities.  For example, if a dataset contains a list of countries, adding the population of those countries as an attribute enables per-capita analyses. 

Traditionally, crafting new attributes and augmenting a dataset with them is done prior to any visual analysis.  Incorporating this process into visual analytics workflows can benefit both the augmentation process as well as the underlying tasks of the system in several ways.  First, the need for an additional attribute may only arise based on insights generated during analysis; if data augmentation is embedded in the system, users can toggle back and forth between analysis and augmentation.  Second, the process of discovering data and crafting a new attribute is an analytical task in its own right, and typically consists of complex querying that can be abstracted away by a visual interface.  Lastly, user-driven curation of the attributes of the dataset can serve as an additional medium of communication between user and system: the user can communicate domain knowledge by adding new attributes, and can likewise learn from the attributes that are discovered by the system.

However, there are many complications in integrating data augmentation into visual analytics systems.  Finding external data and matching it to the entites in the user's dataset is nontrivial.  Once connected to external data, it can be difficult to determine which data is relevant or useful to the user's analysis.  Some attributes may require the aggregation of multiple pieces of external data and can require joining through several intermediate entities.  These hurdles should be abstracted away so that the user is not required to be a database expert.  

Large scale information repositories provide the potential for interactive data augmentation because they put the potential data at the user's fingertips. However, in order to use such repositories a number of challenges must be addressed. If the repository is in the form of a data lake (a large collection of tables, e.g. WikiTables\cite{bhagavatula2013methodswikitable}), it can be difficult to determine which tables are relevant and joinable to the user's dataset. Tabular data can often be fragmented as well: for example, if a user is foraging for data on water usage across the United States, each municipality might be responsible for publishing its own data, and the formats may not align, resulting in sparse, error-prone joins.  Joining together multiple tables can also result in ambiguities over the type of join (left, right, outer, inner) that are difficult to resolve, and limit the types of attributes that can be constructed without additional data munging.  In addition, plenty of external data may be irrelevant or even harmful to the analysis tasks at hand.  
While there have been great strides in addressing some of the technical issues in using data lakes (see section~\ref{sec:related_work:data_augmentation}), little work has been done to allow a user to explore potential attributes and construct new ones.  Knowledge graphs offer an alternative to data lakes because they simplify entity matching and joining ambiguities due to their graph format.  They bring different challenges, however, in scalability and complexity, and there is still a dearth of tools available to facilitate foraging over knowledge graphs.

In this paper, we present \sys, a visual analytics system for Columnar data Augmentation through Visual Analytics.  Carefully crafted attributes are synthesized by running queries over knowledge graphs, as illustrated in Figure~\ref{fig:teaser}, before being added to a dataset as additional columns for downstream analysis.  First, each row of a dataset is mapped to an entity in a knowledge graph. Then, to help the user identify potential information of interest, \sys explores the local graph neighborhood about entities in the dataset to determine commonly held attributes.  Using visualizations of data quality, distribution, and the local topology of the knowledge graph, \sys guides the user through crafting additional attributes of data without any programming or explicit querying.  Users can express complicated operations over the knowledge graph by interacting with these visualizations. In sum, \sys abstracts away the complexity of searching, retrieving, and joining data so that the user can focus on the exploratory task of determining which attributes are relevant to their analysis.

\sys shows the promise of integrating column augmentation as a foraging process within a VA system.  The design process presented in this work in section~\ref{sec:system} can be useful for future visual analytics tools to adopt columnar data augmentation as part of their workflow.  We claim that a resulting VA system can make the exploration and discovery of additional attributes interactive, and let users construct complex attributes without programming.  Users are able to discover new attributes as well as create attributes they had in mind.  The construction of the augmented dataset can improve the outcome of downstream analytical tasks, such as insight generation,  predictive modeling, or any other number of tasks that rely on the presence of a robust dataset.  

We offer evidence of these claims by describing two usage scenarios for insight generation and predictive modeling.  In the first usage scenario, we demonstrate how a domain scientist's analysis can be enriched through cycles of in-situ column augmentation and analysis.  And in the second usage scenario, we show how a user can craft attributes that result in significantly more accurate machine learning models.  We also conduct a preliminary user study on two datasets to assess the usability of \sys.  The results confirm that users are able to both discover and craft relevant attributes easily and quickly.

In this work, we present the following contributions:
\begin{itemize}[topsep=3pt, itemsep=0pt,parsep=3pt]
    \item We present a visual analytics system, \sys, for exploratory data augmentation using knowledge graphs.  We also describe the design process of using visualization as a medium for query construction and knowledge graph exploration. 
    \item We provide usage scenarios of \sys being applied to both insight generation and predictive modeling to demonstrate the generality of our approach.    
    \item We conduct a preliminary user study to assess the usability of our system in joining semantically meaningful external data across two different tasks, offering validation of our design.
\end{itemize}

\vspace{-3mm}

\section{Background and Related Work}\label{sec:related_work}

\subsection{Knowledge Graphs}
Our system uses knowledge graphs as its knowledge representation.  Relational databases are one of the most popular formats to store data because they can be implemented and queried effectively and efficiently.
However, recently knowledge graphs have gained popularity for storing very large datasets because their entity-based model is conducive to how humans think about data.
Knowledge graphs represent data entities as nodes in a graph and the edges between those nodes are relationships between those entities.
The term \textit{knowledge graph} has been loosely used to describe collections of information.
Several different definitions have been offered in recent years~\cite{hogan2020knowledge}.
In practice the term has been used interchangeably with \textit{knowledge base} or \textit{ontology}\cite{ehrlinger2016towards}.
Structuring entities in the form of an ontology is also used to organize conceptual spaces, for example for the evaluation of visual analytics systems\cite{chen2019ontological} and visualizations that support machine learning tasks\cite{sacha2018vis4ml}.

The current popularity of knowledge graphs to store information can be traced back to 2012, when Google introduced its \emph{Knowledge Graph}.
Google's Knowledge Graph is a repository that retrieves facts about entities in search terms on their search results pages\cite{googleknowledgegraph2012, Dong2014}, and is used to power the Google Assistant\cite{googleAssistant} and Google Home voice queries\cite{googleHome}. Recent work has demonstrated that the approach can generalize to other artificial intelligence tasks such as image captioning and conversational agents\cite{huang2019knowledge,zheng2018question,hixon2015learning}.

Since 2012, Wikipedia and other Wikimedia projects have collected information into the \textit{Wikidata} knowledge graph.
It contains broad information about tens of millions of entities.
Other large public repositories exist, from DBPedia\cite{auer2007dbpedia}, a collection of data resources, to the domain-specific universal protein knowledgebase \textit{UniProt}\cite{uniprot2017uniprot} and the linguistic resource \textit{WordNet}\cite{Miller1995}.
Many of these graphs are part of the Linked Open Data Cloud\cite{linkedopendata2020}, a knowledge graph that contains information about other knowledge graphs.
All these graphs contain a tremendous wealth of structured data that can be accessed programmatically with little to no data munging.

Knowledge graphs have been become popular in many research areas such as information retrieval, Natural Language Processing (NLP), and text analysis. For example, a number of approaches (e.g.,~\cite{Dalton2014, liu2015, Rao2013, Raviv2016}) and tools (e.g.,~\cite{Ferragina2010, Gabrilovich2013}) focus on text-centric information retrieval using entity links in underlying knowledge graphs. Examples of applications include the discovery of emerging entities~\cite{Hoffart2014}, the structuring of contents into topics~\cite{Balasubramanian2010}, extracting relations~\cite{Mintz2009}, semantic search~\cite{Berant2013, Krishnan2018, Nanni2018}, or predicting the missing relations between entities~\cite{Dettmers2018, Yang2015}. Researchers in NLP and text analysis fields have also utilized knowledge graphs from different perspectives, e.g., for embedding entities and words into a continuous vector space~\cite{Wang2014}, for extracting relation facts form text~\cite{Riedel2010, Hoffmann2011}, for the generation of questions and answers pairs automatically~\cite{Reddy2017}, and for parsing and interpretation of user natural language semantically~\cite{Hakkani2014}.

In our system, \sys, users can tap knowledge repositories as a data source to augment a dataset through interactive visual support. Knowledge graphs have the advantage of providing clean, curated data which means that the need for data cleaning and matching is greatly reduced.
In addition, the entity-focused way in which knowledge graphs store information can make it easier to think about the relationship of data objects and thus helps users guide the augmentation process.  



\subsection{Data Augmentation}
\label{sec:related_work:data_augmentation}
In the machine learning community, the goal of data augmentation is to expand a training dataset so that as much of the phenomenon being modeled is present as possible.
This is done in two ways: adding objects (rows) to the training data, and adding new attributes (columns) for the objects.
For example, in image datasets, adding new objects in the form of slightly modified versions of the training objects (i.e. rotating, adding noise, cropping, modifying color) can improve a machine learning model's sensitivity to noise in the data\cite{krizhevsky2012imagenet, Shorten2019, perez2017effectiveness, Jung2020, Park2020}.

Approaches to add columns to a dataset typically vary in where that data comes from, and whether that data is used in training the model or in model inference.
Feature engineering\cite{kanter2015deep} is a common method to derive new attributes from existing columns by applying operations to them, such as the difference of two columns.  However, it is limited in that it can only express data that is in the scope of existing attributes.
Knowledge graphs have recently been used to incorporate world knowledge into machine learning models~\cite{Song2017} for a range of models, including text processing\cite{Annervaz2018}, image classification\cite{Marino2017}, and machine translation\cite{Moussallem2019}, but most work incorporates knowledge graphs into the last step of the machine learning pipeline, inference, to gather facts, rather than augmenting an entire dataset.
The Python library RDFFrames\cite{Mohamed2020} helps extract data from knowledge graphs to improve machine learning training.
It allows practitioners to effectively express queries to knowledge graphs and execute them efficiently.  However, it requires extensive data science and programming experience to use.

Automated augmentation of datasets is of interest to the database and the data science communities.
In the database community, the goal of data augmentation can manifest in many ways. 
For example, before augmentation can take place, the first challenge is to find datasets that are suitable for joining. 
In this scenario, sometimes referred to as a ``data lake''\cite{Miller2018}, the data joining system assumes that there is a finite number of candidate datasets, and the task is to identify which of them can be joined with a user's base dataset.
Systems such as Google Goods\cite{halevy2016goods}, Infogather\cite{Yakout2012}, Octopus\cite{cafarella2009data}, Aurum\cite{fernandez2018aurum}, and work by Sarma et al.\cite{das2012finding} examine the attributes of the candidate datasets and learn the relationship between those attributes and the attributes of the input data.

One challenge of automated data augmentation that is an active area of research is entity matching. 
Entity matching refers to finding the same entity in different datasets, often using machine learning techniques. Once identified, the entity can be removed (for deduplication) or merged (for augmentation). Examples of entity matching systems include Magellan\cite{konda2016magellan}, Nadeef\cite{dallachiesa2013nadeef}, Autojoin\cite{zhu2017auto}, and work by Mudgal et al.\cite{mudgal2018deep}.

What the data joining and entity matching approaches have in common is that the systems assume little knowledge about the data. The challenge is therefore to identify the commonalities between the datasets (e.g. schema matching) to determine if the datasets or the entities within are related and therefore joinable.
In \sys, we take a different approach.
Instead of assuming little to no knowledge about the candidate data, we use a knowledge graph as the source of ``candidate data.'' 
Although knowledge graphs can be loosely structured, they contain well-defined relations between entities. 
This property of knowledge graphs allows \sys find relevant data for augmentation with certainty and ease.
While automated approaches can discover joinable attributes for a given dataset, they lack semantic knowledge of expert users.
This may result in a large number of attributes that are irrelevant to the analysis problem being added to the dataset, hindering the users in further analyzing and gaining insights from the dataset.
To make use of their semantic knowledge to choose relevant attributes, users may benefit from an exploratory process facilitated by visual analytics.  Beyond the selection of relevant attributes, users may benefit from visual explanations of complex queries.  For example, in transitive relationships that go through multiple entities, joins require multiple operations to be specified for aggregating a collection into a single value.
\sys overcomes these issues by involving a user in the process.  Attributes are only added to the dataset if the user decides that they are helpful for their analysis.  

\subsection{Visual Analytics Approaches}

Curating, improving, and augmenting an existing dataset has been a popular research topic in visual analytics.  This can be done to aid interpretation and sensemaking\cite{cramer2017impact} during analysis.  Other visual interactive approaches allow users to query and integrate query data from heterogeneous web or local sources.
This includes helping users extract information from textual web sources\cite{Hoeber2017}, query knowledge graphs\cite{Hoefler2014}, query large metadata-rich heterogenous data stores~\cite{Smith2006}, identify relations at attribute levels in mixed data sets~\cite{Jurgen2014}, analyse associated categories in large categorial data tables~\cite{ Alsallakh2012}, or automatically create visual representations of information stored in knowledge graphs. VAiRoma helps users extract and combine information from Wikipedia articles to create visualizations that provide insight into historical events\cite{Cho2015}.
Vispedia\cite{Chan2008} lets users interactively collect and integrate data from Wikipedia tables to create visualizations and answer analysis questions.
Atlasify enables users to relate the query concept, corresponds to a Wikipedia article, to spatial entity in the underlying reference system, e.g., countries, political figure~\cite{Hecht2012}.
All of these approaches help users access information stored in different forms, but compared to \sys do not support the process of augmenting an existing dataset with additional data.

\section{Knowledge Graphs}

\label{sec:knowledge_graphs}

We define our usage of the term, knowledge graph, in this section, and describe the sufficient characteristics of a knowledge graph that we assume for our system. We assume that knowledge graphs contain entities as nodes and express attributes of those entities as edges, connecting the source entity to either another entity (to express a type of relationship) or to a literal (such as a string, number or datetime).  The number of edges connected to a node can be a proxy for the number of attributes that are held by that entity.  The flexibility of this structure allows for complex relationships to be expressed between entities, including one-to-one (i.e. a country has one head of state), one-to-many (a country is composed of multiple municipalities), and many-to-many (countries share borders with other countries non-exclusively).


\sys requires that the connected knowledge graph has a data retrieval endpoint that can respond to simple queries about entities and their neighbors.  The connected graph must also have some sort of service to map from the values in the dataset uploaded to \sys to the entities on that knowledge graph.  For example, if a dataset with U.S. states is uploaded, there must be an existing service that maps from ``New York'' or ``NY'' to the entity in the knowledge graph corresponding to that state.  These two requirements allow \sys to connect a user's data to the knowledge graph and to retrieve relevant data for the user to forage from.


In the experiments in this paper, we connect \sys to Wikidata.  It meets both requirements listed above: it responds to the SPARQL query language so gathering neighborhood data about entities is simple, and it has a service \verb wbsearchentities \hspace{1mm}to map strings to entities on the graph.  But it also has some extra features that we take advantage of.  Wikidata contains a broad set of information making it easy to find related attributes for many different kinds of datasets. It also has additional metadata about the data in its graph including data-type information and human readable descriptions and labels for each node and edge in the graph.  Lastly, due to its popularity, there is a large amount of documentation and guidance on constructing complex queries.  

In future work, we hope to allow users to connect \sys to multiple knowledge graphs.  To limit scope in this work, we focused only on Wikidata because of its high standards for data quality.  In addition, it contains general knowledge that made it an applicable information repository for a number of different datasets and usage scenarios.  This also made it easier to recruit users for the study described in section~\ref{sec:user_study}.  While different knowledge graphs bring different challenges, particularly around data sparsity (see the discussion in section~\ref{sec:sparsity}), Wikidata provided a broad enough testbed to inform the design goals and interactions supported by \sys.



\section{Tasks and Goals}



\sys is developed as part of the DARPA Data-Driven Discovery of Models (D3M) program whose goal is to develop software infrastructure and algorithms to make automated machine learning accessible to general data scientists. Inspired by the observation that both the use of exploratory visualization and the use of advanced machine learning are fundamentally limited by the input data, \sys was developed to help a data scientist craft better predictive models by foraging for additional attributes to add to their dataset.

We conducted interview sessions with four teams within the DARPA D3M program developing applications for data exploration and predictive modeling. The goal of the interview sessions was to better understand how a tool like \sys can help the user to perform data augmentation for the purpose of improving machine learning model performance and accuracy.
Each of the interviewed teams was shown an early implementation of \sys that was able to search for related attributes and return a list of them, but without any visualizations. The participants were then asked about what additional system features and interactions would facilitate the discovery of the types of data that they were interested in for their applications.  
\begin{figure*}[thb]
\centering
\includegraphics[width=0.92\textwidth]{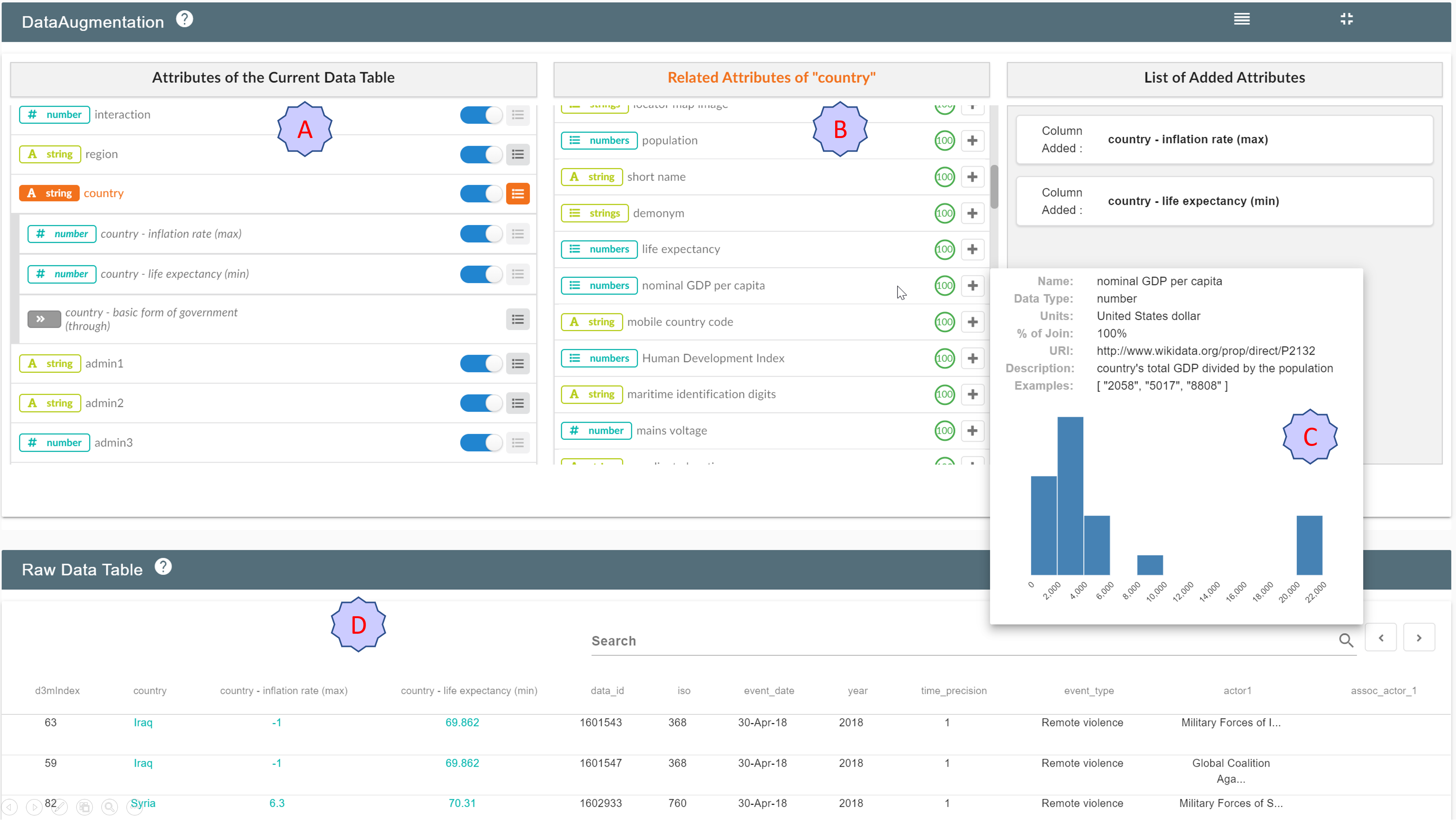}
\caption{The user interface of \sys, showing an analysis session on the ACLED dataset.  \textbf{A} shows the initial list of attributes in the table as well as augmented attributes, like the maximum inflation rate recorded for each country.  The list of attributes related to the \textit{Country} column is shown in  \textbf{B}.  Each related attribute is adorned with a donut chart showing the estimated join quality.   Additional details about any attribute is available on demand in a popup, as seen in \textbf{C}.  As attributes are added to the dataset during the analysis session, a preview of the table is updated as seen in \textbf{D}.}

    \vspace{-3mm}

    \label{fig:system_overview}
\end{figure*}


\subsection{Task Analysis}

We distilled a list of tasks from the interviews that would enable a user to meaningfully augment their dataset with additional attributes.

\begin{enumerate}[label=\textbf{T\arabic*:}, topsep=0pt, itemsep=0pt,parsep=3pt]
    \item \textbf{View a list of joinable attributes for any existing attribute in the dataset.}  Revealing the list of potential attributes that can be joined with a particular column in the dataset helps the user determine what external data is available for augmentation.  It is a key aspect of the exploration process in information foraging because it may reveal data that the user wasn't aware of.
    \item \textbf{Analyze the properties of possibly related attributes \textit{before} the join occurs:}
    Before joining a new attribute into a dataset, users will need to gather information about the new attribute to gauge the new attribute's impact on the quality of the dataset.
    For this reason, users should be provided with as much information as possible about the potentially joinable attributes before the join.
    This might include metadata, examples of the attribute, and any available information about the general distribution of the new data.  In addition, it is important to communicate whether there will be any missing values in the joined attribute.  
    
    \item \textbf{Specify aggregations:}
    The relationships between entities in the user's dataset and entities in the external data source are varied - they could be one-to-one, one-to-many, or many-to-many.
    In order to fit information encoded as plural relationships into a single row of data, aggregations must be specified.
    For example, a user might want to add the populations of a list of states into the their data. However, because there could be multiple measures of the state's population over the years, the user would need to perform an aggregation function over these results, such as \textsc{minimum}, \textsc{maximum}, or \textsc{average} depending on their analysis need.
    
    \item \textbf{Connect through to additional data:}
    Attributes relevant to the user's analysis goals may not always be directly encoded as a property of an entity that exists on the initial dataset.
    For example, a user may want to augment a dataset of countries with the population of each country's capital.
    This information may not be directly accessible as a property of each country in the data repository.
    Instead, population size is linked to the entity that represents the capital, which in turn is linked to the country.
    To support these cases, users should be supported to join \textit{through} intermediate attributes giving them fine-tuned control over how joins pass through multiple relationships.
\end{enumerate}

\subsection{Design Goals}

Based on this task analysis, we iterated through several sketches to come up with a set of design goals.  In the subsequent section, we describe a system with visualizations and interactions that address these goals.  However, this list of design goals should prove to have a broader impact than \sys alone - they can inform how data augmentation can be integrated into other visual analytics tools.

\begin{enumerate}[label=\textbf{G\arabic*:}, topsep=0pt, itemsep=0pt,parsep=3pt]
    \item \textbf{Attributes as first-class citizen.}  Because external data is found based on attributes in the original data (\textbf{T1}), the list of attributes is the most important data to encode.  In early sketches, we used a data table to show the user what the data looked like, and allowed users to click on column headers to view new attributes and view joined data in enriched cells, similar to the \textit{ETable} from Kahng et al~\cite{kahng_2016}.  We ultimately decided that a table encoding dedicated too much space to the table formatting and data found in the table cells.  Instead, we settled on a primary view that listed all attributes of the dataset.  By viewing the list of attributes, users are able to get a sense of what attributes might still be needed for their analysis.  This list can update as new attributes are foraged.  Additional details about each attribute, such as summary statistics and examples, can be provided on demand.
    \item \textbf{Visual exploration of joinable data:} In order to discover interesting joinable data (\textbf{T2}) and specify aggregations on that data to create new attributes (\textbf{T3}), users should be able to explore the set of joinable data.  Visual cues can help in this exploration, by encoding meaningful information about external data like the join quality or the data distribution.  As the dataset evolves, users should be given previews of the data they have added to enable iterations of exploration.
    \item \textbf{Provide visual examples for complex queries.} In practice, building complex queries that stepped through intermediate information on the knowledge graph becomes very complicated because it requires multiple aggregations over the joined data.  For these types of complex queries, visualizing the knowledge graph about one example in the data can help a user understand how they are combining and shaping the data.
\end{enumerate}

\section{\sys: A Visual Analytics System}

In this section, we describe the function and design of \sys.  The general workflow of \sys starts when a user uploads a tabular dataset.  The user can then search for related attributes of any column in their dataset containing entities that can be found in Wikidata.  They can augment their dataset with these related attributes by specifying aggregations.  After several iterations, if the user is satisfied with the augmented dataset, they can export their data for further analysis in other visual analytics tools.  

In this section, we first describe the design of each component of the interface and their interactions.  Then, we explain how these user interactions are translated into queries that can be executed over a knowledge graph.





\subsection{User Interface}

Figure~\ref{fig:system_overview}, \sys has the following main interface components:

\begin{figure}[t]
    \centering
    \includegraphics[width=\linewidth]{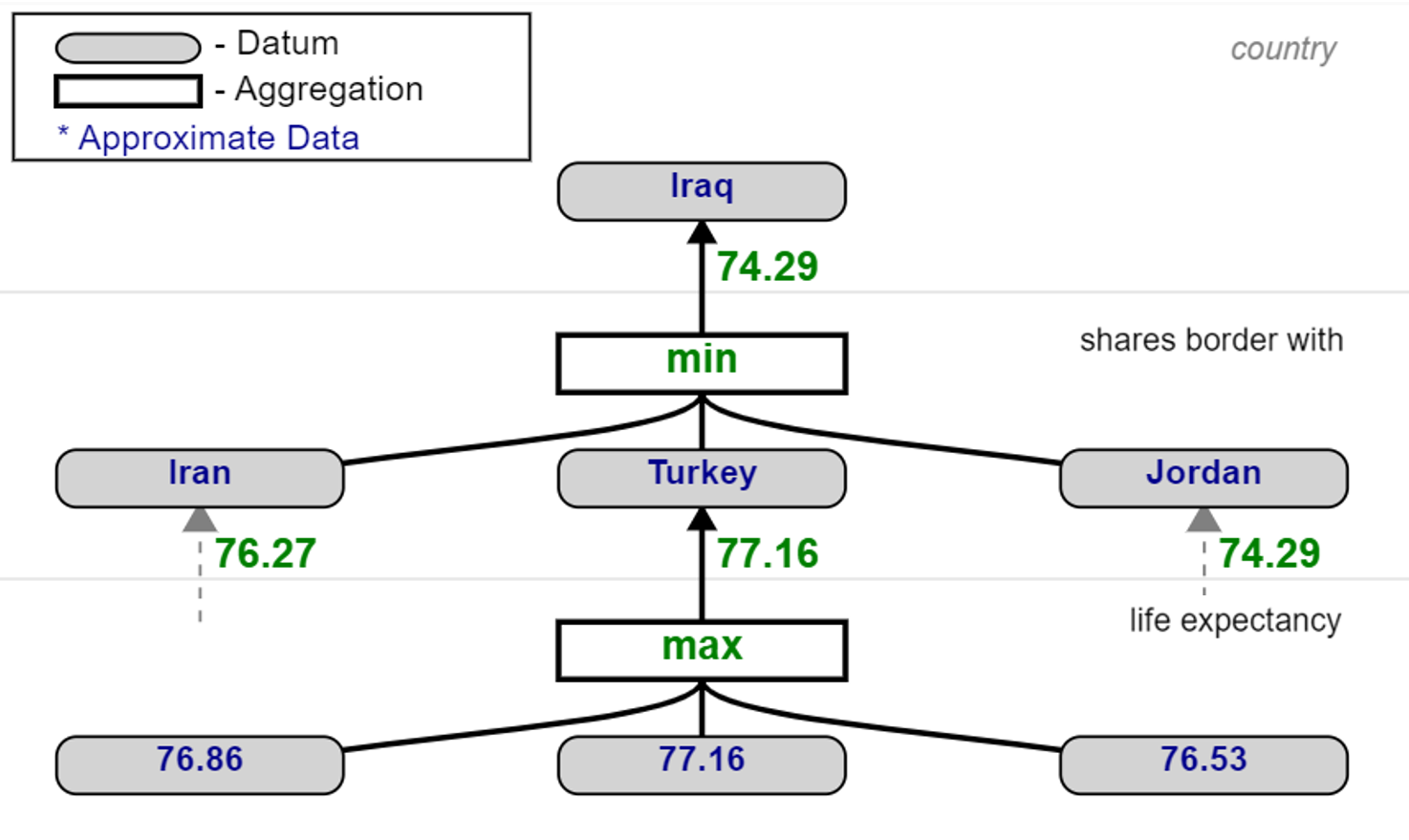}
    \vspace{-4mm}
    \caption{When a through join is specified, a dialog window pops up to allow the user to specify aggregations at each level of the join.  To help them specify that query to the system, \sys shows them an example of the neighborhood in the knowledge graph that is queried for a single row of the table.  In this image, the user would like to add the lowest life expectancy of any neighboring countries because they believe this will help their analysis of conflict.  And for each country, several values are available for life expectancy, so the user selects the ``max'' operator.  All six neighbors for Iraq, but only three are shown for simplicity of the illustration.  }
    \vspace{-3mm}

    \label{fig:through_join}
\end{figure}

 \vspace{3pt} \noindent \textbf{Column View:} This view shows the attributes and their data types (\textbf{G1}, see Figure~\ref{fig:system_overview}-A). Users can view details on demand about each attribute, including the data distribution of that attribute shown as a histogram chart, several example values, the type of unit (i.e. kilogram or mile, if applicable) (see Figure~\ref{fig:system_overview}-C).
 
 \vspace{3pt} \noindent \textbf{Related Attributes:} A user can search for related attributes of any string attribute in the Column View.  \sys shows the related attributes discovered from the knowledge graph next to the list of attributes of the current table, as seen in Figure~\ref{fig:system_overview}-B. 
 An estimation of the join quality of each related attribute is encoded visually with a donut chart, and details on demand are also provided for each attribute, including distribution of values.
 This information can help the user identify whether a related attribute has the potential to help their analysis \textit{before} joining it to their data (\textbf{G2}).  Both the estimated probability and the estimated attribute distribution are calculated from random sample of possible joinable data, since calculating the true values would require completing the full join for every row in the dataset.  For more details on implementation techniques, see section~\ref{sec:technique}.

 \vspace{3pt} \noindent \textbf{Adding Attributes:} To add any attribute, users click the plus symbol, and \sys will show a drop-down menu revealing various options for join operations. The available join operations correspond to the data type of the relationship.  If the relationship is one-to-one, i.e. a country has a single head of government, then a single value will be retrieved for each row of the dataset.  If the relationship is one-to-many or many-to-many, an aggregation must be specified.  For example, if the attribute is a collection of numbers, such as set of populations recorded for a country, the user can select numerical aggregations such as \textit{count}, \textit{mean}, \textit{max}, \textit{min}, \textit{sum}, or \textit{variance}.  If the attribute is a datetime, the user can select from \textit{count}, \textit{max}, or \textit{min}.  If the attribute is a string, the user can select either \textit{count} or \textit{through}, a type of placeholder operation that let's the user step through this attribute to form complex multi-hop joins (more details are provided below).  In addition to these join operations, for any type of data, the user can select to randomly sample from the collection.  This operation is useful if an attribute is generally expected to have only one value per entity, but due to data quality issues, some rows might have it recorded twice, like date of birth.  
 
 Extracting information for the whole dataset from the knowledge graph is too computationally demanding to be performed at interactive rates.  Gathering the data from the knowledge graph requires the retrieval of many different parts of the knowledge graph in contrast to the retrieval of a single column from a relational database.  \sys mitigates much of that time cost by only joining attributes for the top 10 rows of the dataset.  Users are still able to glance at the dataset preview, as seen in Figure~\ref{fig:system_overview}-D, and get an idea of the data that is being joined to the dataset (\textbf{G2}).  This lets the user explore the available attributes rapidly.  When the exploration phase is completed, the full join can be executed.  
 
 \vspace{3pt} \noindent \textbf{Through Joins:} In some cases, the user may want to join to data through an intermediate attribute.  If the intermediate attribute has a one-to-one relationship with the rows of the original dataset, i.e. a country has one head of government, then the user can simply join the intermediate attribute first, and then use that as their starting point to seek more related attributes.  If the intermediate attribute is a one-to-many relationship, then the user must specify multiple levels of aggregation.  We call this type of aggregation a \textit{``multi-hop aggregation,''} since it requires aggregating data across multiple hops on the knowledge graph.  
 
 As an example, suppose the user wants to determine if a country has a lower life expectancy than its neighbors.  Consider how that would be calculated for a single country, Iraq.  To gather the required data from the knowledge graph, the user must first join to all bordering countries, then connect through them to reach the life expectancies of those countries.  Once all data is selected, aggregations must be specified to produce the desired value.  In Wikidata, countries have multiple life expectancies recorded, so the user has to specify that they want the maximum life expectancy recorded.  That value is then calculated for each bordering country (i.e. Iran, Turkey, Jordan, etc.).  Then, they have to specify that they want the minimum among all bordering countries.  Expressing this type of query is complicated enough to explain for a single value, let alone for an entire column.  
 
 In \sys, the user is shown a simplified illustration of the topology of the knowledge graph to assist them in understanding the aggregations that the user must choose, as seen in Figure~\ref{fig:through_join}.  For a given row of the dataset, up to three values of the intermediate attribute and the target attribute are sampled from the knowledge graph.  The system then displays these values on a graph, representing a small slice of the knowledge graph.  Users can then dynamically set the aggregations (i.e. \textit{count}, \textit{mean}) that occur at each level of the join graph, and see the resulting value that would get joined for that row.  By viewing the values that will get \textit{passed through} the knowledge graph to ultimately calculate the attribute for one row of their data, users get reinforcement that the complex query they are building in \sys matches the query they are building in their head (\textbf{G3}).

\subsection{Backend Implementation}

\label{sec:technique}

Here, we describe how our visual analytics system augments a dataset by translating the tasks supported by the user interface described above into valid queries over a knowledge graph.  \sys gathers data from the connected knowledge graph in two different subroutines.  In all examples in the paper, \sys connects to Wikidata using the SPARQL query language.\footnote{Example queries generated by the system for both subroutines are available in the supplemental material.}



\vspace{3pt} \noindent \textbf{Finding Related Attributes:} This subroutine receives a column of data as input and returns a list of attributes along with assorted metadata.   Consider a dataset with a Country column, i.e. ``Germany,'' ``France,'' ``Austria,'' etc..
Unlike a column in a SQL table, where entities and their relations might be expected to adhere to a schema, a list of countries in a knowledge graph might not share the same set of attributes. For example, in Wikidata, some Western countries, such as Germany, can have more than a thousand attributes, while other countries have only a fraction of that.  
In order to provide a user with a consistent set of attributes for these entities, we first need to find their commonalities.
As such, the primary goal of this subroutine is to return the list of attributes that are available on as many of the entities as possible.  

Checking the related attributes of each entity to gather this list can take prohibitively long, especially if the input data contains thousands to millions of rows.  As an optimization step to make this subroutine support a user's interactive analysis, we employ a sampling-based technique to reduce computation time. With this optimization, some subset of the dataset is randomly sampled; in our experiments we have found that 20-30 rows are sufficient to differentiate high-quality attributes from poor-quality ones. Each sampled row is mapped to an entity on the knowledge graph, and a list of related attributes for that row is retrieved.  Then, the lists for all the rows are compared, and the 50 attributes that appear on the most lists are returned.  In this query, we also gather all details of the attributes that are displayed on demand, such as the data distribution of that attribute shown as a histogram chart, several example values, and the type of unit (i.e. kilogram or mile, if applicable).  For the types of exploration done in the user study and in examples given in this paper, retrieving the list of related attributes from Wikidata typically takes 2-5 seconds.

    


\begin{figure}[ht]
    \centering
    \includegraphics[width=0.45\textwidth]{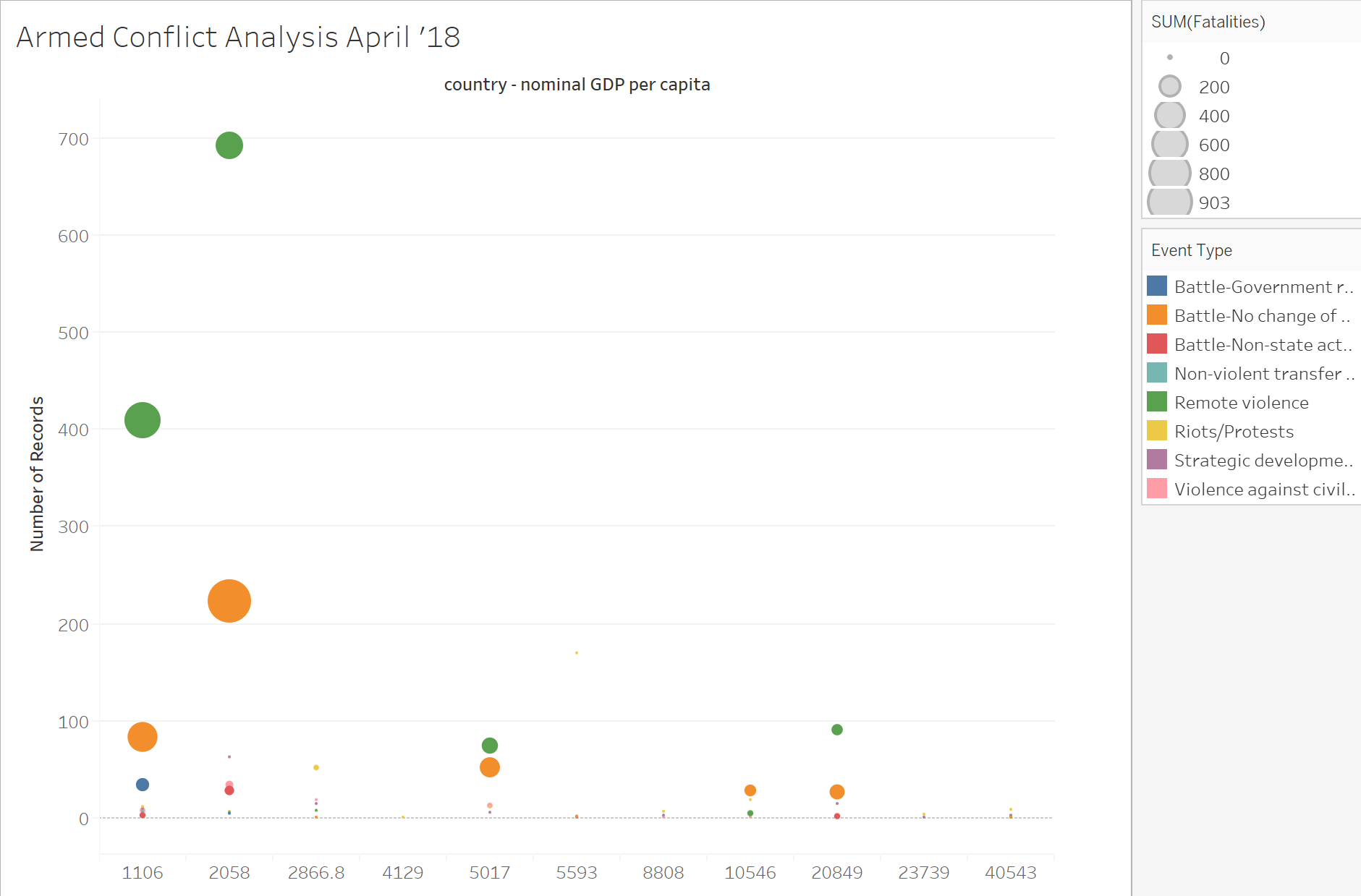}
        \vspace{-3mm}

    \caption{Visualizing the number of records in the ACLED dataset in April 2018 by per capita GDP, a field extracted from Wikidata.  Magnitude of marks encodes number of fatalities, while color encodes event type.    Visualization generated with Tableau. }
        \vspace{-3mm}
    \label{fig:tableau_use_case}
\end{figure}

\vspace{3pt} \noindent \textbf{Materializing Joins:} This subroutine takes in join instructions and returns an augmented dataset with an additional column.  The join instructions specify the path taken through the knowledge graph, along with any aggregation functions, such as ``count'' or ``max.''  \sys builds a SPARQL query to crawl the knowledge graph to retrieve the desired data (see Fig~\ref{fig:teaser}).  The time to retrieve data scales linearly with the number of rows and is limited by the concurrency limits on the knowledge graph API. While the user is interactively exploring attributes, \sys will only materialize the join for the top 10 rows of the data table to provide the user with a preview of the table in real time (see Fig~\ref{fig:system_overview}-D).  When the foraging is done, all joins are materialized for the entire dataset (rather than just a sample of rows) in the order in which they were constructed by the user, which takes around 10-15 seconds on the datasets explored in the user study in section~\ref{sec:user_study}.




\sys is implemented with a VueJS frontend web application and a NodeJS backend to handle all asynchronous calls.  Queries are sent concurrently whenever possible.  Source code and installation instructions are available at \url{https://tuftsvalt.github.io/snowcat/}.




\vspace{-0.5mm}

\label{sec:system}

\section{Usage Scenario 1: Conflict Data}


To demonstrate the value of column augmentation  for insight generation, we present a usage scenario in which \sys is integrated into a typical workflow for the popular visual analytics tool, Tableau.   Suppose a political scientist wants to study the factors that lead to armed conflict.  They download the Armed Conflict Location \& Event Data Project (ACLED) dataset for April 2018, which contains 2,279 records of armed conflicts across many different countries and dates\cite{raleigh2010introducing}.  Initially, they load the data into Tableau to analyze relationships between countries, event types, and fatalities stemming from armed conflict.  They see no clear pattern that emerges as they find it difficult to visually group countries without more data.


They load their dataset into \sys to look for additional attributes that may be relevant to event types.  When the system is first loaded, the user sees a list of the attributes in the uploaded table in Figure~\ref{fig:system_overview}-A.  They also can see the first five rows of the data below in Figure~\ref{fig:system_overview}-D.  

The political scientist believes there might be economic metrics that effect the types and severity of conflicts, so they click the \textit{related attributes} button next to ``Country'' in Figure~\ref{fig:system_overview}-A.  The system returns a list of related attributes found by scraping Wikidata, as seen in Figure~\ref{fig:system_overview}-B (\textbf{T1}).  The user scans through the list of related attributes, looking for information about the countries' economies.  They see that Wikidata holds the \textit{nominal GDP per capita} for each country.  
The user adds this attribute to the dataset.  While looking for economic data, the user also notices additional interesting attributes for analysis and joins them for future analysis, including the max \textit{inflation rate}, minimum \textit{life expectancy}, the \textit{type of government}, and the mean \textit{Human Development Index} (\textbf{T3}).

The user loads this next iteration of the dataset into Tableau, and generates various visualizations including the event types and fatality numbers with their newly-discovered attributes.  As can be seen in Figure~\ref{fig:tableau_use_case}, grouping event types by nominal GDP per capita reveals a trend that certain event types, like \textit{Remote violence} and \textit{Battle - No change of territory}, appear to have more severe fatalities in countries that have lower nominal GDP per capita.  In contrast, other event types like \textit{Battle - Government regains territory} and \textit{Battle - Non state actor overtakes territory} don't seem to be very different between high nominal GDP per capita countries and low ones.  The additional visual analysis enabled by the added attribute, nominal GDP per capita, has led the user to new hypotheses that can then be further explored.  This usage scenario demonstrates how the integration of column augmentation into a traditional insight generation pipeline can unlock new analyses in-situ.

\section{Usage Scenario 2: Modeling Poverty}

\label{sec:use_cases_ml}

In our second usage scenario, we demonstrate the value of \sys for a separate type of visual analytics task - the generation of a predictive model.  Many visual analytics systems have been constructed to enable domain experts to interact with and steer the generation of machine learning models on their data.  Augmenting a dataset with new attributes is one way in which a user can imbue domain knowledge into the modeling process.  They may know that certain attributes can help the prediction while others may mislead.  While it may seem unusual to augment a training set with additional attributes, the additional attributes found on the knowledge graph can likewise be added to any data that the resulting model is asked to predict on at test time when the model is deployed.

To demonstrate the use of CAVA in the model building process, we modify \sys to be able to train a predictive model at any point in the data augmentation process. 
Users can instruct the system to build a predictive model with the current version of the data.  \sys splits the data into a training and test set in the ratio of 0.8:0.2 and trains a random forest regressor.  For each model, \sys reports the \textit{R Squared} score on the training and test sets, as well as the feature importance scores of the five most important features.  This functionality was designed to be representative the iterative workflow of model builders, and to examine the ways \sys could help in-situ model building.

Consider \userML is a public policy analyst who seeks to build a regression model on a U.S. Census Bureau dataset containing data about poverty in different counties around the U.S.\cite{poverty2020}.
The dataset contains $3136$ rows, and $6$ attributes--the dataset index, \textit{FIPS}, \textit{State}, \textit{County}, \textit{RUCCode}, and the number of residents living in the county under the poverty line.
\userML seeks to augment this data by adding meaningful columns to accurately predict the \textit{Poverty}. 
When \userML first loads the data in \sys, an initial model is trained with an \textit{R Squared} score of $0.577$ on training and $-1.801$ on test set respectively; indicating that the model badly overfitted the training data. 
\userML seeks to improve the regression model's performance further by augmenting the base data using \sys's workflow and visual interface.
\userML searches for related attributes of the variable \textit{State} from the Column View. In response, \sys shows a list of related attributes that \userML may consider to add to the data. 
From these attributes they click on a set of interesting attributes such as \textit{shares border with}, \textit{life expectancy}, \textit{inflation rate}, etc. to see the distribution of values, and its metadata as a text view. 

From these set of attributes \userML thinks that the attributes \textit{Inflation Rate} and \textit{nominal GDP per capita} are good attributes to predict \textit{Poverty} and thus decides to add them to the data using the join operation \textit{Mean}. They notice that the newly added column is displayed on the Table View. However looking at a few rows of the table \userML finds quite a few cells that are empty indicating that the joined data may have missing values. 
Nonetheless, \userML clicks a button to construct a new regression model using the augmented data.
\userML notices that the \textit{R Squared} score marginally improves (new score: $0.586$ and $-1.810$ on training and test set respectively). They hover the mouse over the model metric card to see the list of ``Top $5$'' attributes with their weights utilised in the regression model. 
While \userML expected to see a substantial improvement in the model performance, 
they infer that the marginal improvement is probably due to missing values in the data after the join operation. \looseness=-1

Motivated to improve the model further, \userML searches for columns that may directly help to predict poverty per county. Based on prior knowledge, \userML understands that the \textit{population} of a state may be directly proportional to its poverty, and they add that attribute. 
In the process of searching for other relevant columns, \userML notices the column \textit{Maximum temperature}. Inquisitive to see if a temperature of a state is correlated with its \textit{Poverty} they add it to the table.
After constructing a new regression model, \userML notices that the \textit{R Squared} score changed from $0.586$ to $0.577$ and from $-1.810$ to $0.541$ for the training and test dataset respectively. Happy with the progress so far they decide to remove any column that may not be contributing to the prediction task. First, they remove the columns \textit{RUC Code}, and \textit{nominal GDP per capita} (by triggering the slider on the Column View) and then triggers \sys to construct a new regression model. 
As expected \userML notices that the \textit{R Squared} score did not change. 
Next \userML searches for related attributes of the column \textit{County}.
They explore the set of choices shown in the Column View.
They choose to add  the column \textit{County-population} (by median operation), and \textit{shares border with} (by count operation). \userML constructs a new regression model to see the models' train and test \textit{R Squared} score improved considerably ($0.807$, and $0.746$ respectively). Content with the improvement, they export the model and the data to continue analysis.

In a short time, \userML has constructed a regression model with substantially better R2 score than the model trained on the base data.  They have also gained insight into which attributes are relevant to their modeling problem, which may help their understanding of the resulting machine learning model.

\section{Preliminary User Study}
\label{sec:user_study}
We evaluate \sys in a preliminary user study. The purpose of the evaluation is to validate \sys both in terms of its usability and its effectiveness in helping a user improve a machine learning model through data augmentation. 
Specifically, we hypothesize that:

\begin{itemize}
    \item \textbf{H1}: \sys allows users to accurately join external data given a written description of that data. 
    \item \textbf{H2}: \sys is able to help users discover additional data that can improve the predictive quality of machine learning models trained on the dataset.
\end{itemize}

We recruited $6$ participants (3 Female, 3 Male), between the age of $23-36$. We required each participant to have at least an elementary knowledge of machine learning and data analysis. Due to COVID-19, we were not able to conduct an in-person study. Instead  we conducted an online study using Bluejeans\footnote{https://www.bluejeans.com/}. The participants interacted with \sys on their own computer while sharing their screen.
We provided the participants with a url to our system that was hosted on our local machine that is exposed using the Ngrok remote tunneling software\footnote{https://ngrok.com/}. The study took approximately $50-60$ minutes and we compensated the participants with a $\$10$ Amazon gift card.

\subsection{Study Design}
Before the study we asked participants to fill out a background information questionnaire regarding their name, age, gender, machine learning expertise and various use cases in which they use machine learning.
We began the study by showing the participants a tutorial video of \sys, explaining the workflow, interface GUI elements, and its interaction capabilities that support various join operations. Next we asked the participants to perform three tasks, the first of which was a practice task to ensure that they were 
sufficiently knowledgeable about \sys to perform the experimental trials. We proceeded to the experimental sessions only when we observed that the participants were confident and able to use \sys on their own. In the next two tasks we asked the participants: (1) To add a set of specified attributes to a given dataset. These attributes can be added to the data using various join operations supported by \sys. (2) To freely augment the data such that they can improve the performance metric of a machine learning model (metric being a R Squared Score of a regression model). 
We used the Scikit Learn machine learning library \cite{scikit_ml} to construct \textit{Random Forest Regression} models, in the same manner as described in Section~\ref{sec:use_cases_ml}.
We collected the following data from each experimental trial: (1) \textit{Task competition time}, (2) \textit{Task Accuracy}, (3) \textit{Model performance metric} e.g., R Squared Score, and (4) \textit{User ratings} collected using a post-study Likert-scale questionnaire.

\subsection{Datasets}

For the tutorial video and practice task we used the same U.S. Census Bureau dataset as section~\ref{sec:use_cases_ml}.
For the first task in our experimental session we use the IMDB Movies dataset\cite{imdb2020} containing $500$ movies and $28$ attributes such as 
\textit{Director-name}, \textit{Duration}, \textit{Movie-Title}, \textit{Movie Cast Facebook Likes}, etc. The second task in the experimental session which required users to augment data and construct regression models, used the unemployment rate dataset\cite{unemployment2015}. This dataset contained $1200$ rows where each row corresponded to a county's poverty rate measured in a given month. This dataset contained only $5$ attributes--the dataset index, \textit{Month}, \textit{State}, \textit{County}, and the \textit{Unemployment-Rate} (dependent variable). 

\subsection{Tasks and procedure}
Participants were first asked to complete a practice task.  
During the practice task, their answers were not recorded and their performance was not included in the analysis described below.  
For the practice task, using the Poverty dataset, we asked the participants to add three attributes: (1) the area of the county, (2) the population of the state, and (3) the earliest inception date of any county that each county shared borders with. 
While the first two attributes could be retrieved using a straight-forward \textit{join by value} operation, the third attribute required the \textit{join through} operation to search for the required attribute that was one ``hop'' away in the knowledge graph. For the first experimental task using the IMDB dataset, participants were given descriptions of four attributes to add: (1) the director's country of citizenship, (2) the number of awards received by the director, (3) the number of cast members in each movie, (4) the sum of the number of awards received by all the producers of each movie.  
Note that attributes 3 and 4 required the participants to \textit{join through} several attributes.  

The final experimental task gave participants $10$ minutes 
to augment the Unemployment rate dataset with as many attributes as they'd like that they believed would aid in building an accurate regression model to predict poverty rate.  
At any point in the 10 minutes, participants could instruct \sys to build a regression model with the current dataset to give them insight into whether they were improving the predictive modeling process.  
Building a model took about 10-20 seconds, and participants would be shown the change in R Squared Score for each new model, as well as the  weights of top ``5'' attributes used in the model. 
Random Forest regression models were used because their flexibility, speed, and accuracy met the constraints of the experiment.


\subsection{Data Collection}

After all the tasks were completed, participants were asked to fill out a post-study questionnaire (using Google Forms) that included: (1) Likert-scale\cite{likert1932technique} rating questions on their use of the system, (2) a NASA-TLX\cite{hart1988development} questionnaire to measure task difficulty, and (3) Open-ended descriptive questions asking users about the interface design, the workflow, and other responses related to improve the usability of the system\footnote{Pre-experiment and Post-experiment surveys are attached as supplemental material}. 
The Likert-scale questions asked the participants to rate (in a scale of $1-7$, $1$ being ``strongly disagree'') if they found the system: (1) Easy to learn, (2) Intuitive, and (3) Expressive for data augmentation. 
We used the open-ended descriptive feedback to qualitatively evaluate the usability and the interaction design of \sys for their tasks.
Open-ended prompts included (1) \textit{Describe your thoughts about \sys}, (2) \textit{Describe things you disliked about the system or the workflow. Elaborate on how you think it could have been improved}, and (3) \textit{Describe your strategy or your process to augment the data for Tasks 1 and 2}.
With the consent of the participants,
each session was both video and audio-recorded. We encouraged the participants to verbalize their thoughts following a think-aloud protocol.
Furthermore, to assess if data augmentation using \sys led to any change in the regression models performance, we saved: (1) model metric (i.e., R Squared Score), (2) attribute weights, and (3) predicted values from the model. We also saved user mouse-clicks to analyze the set of attributes the user explored to augment the data.

\begin{figure}[t]
    \centering
    \includegraphics[width=\linewidth]{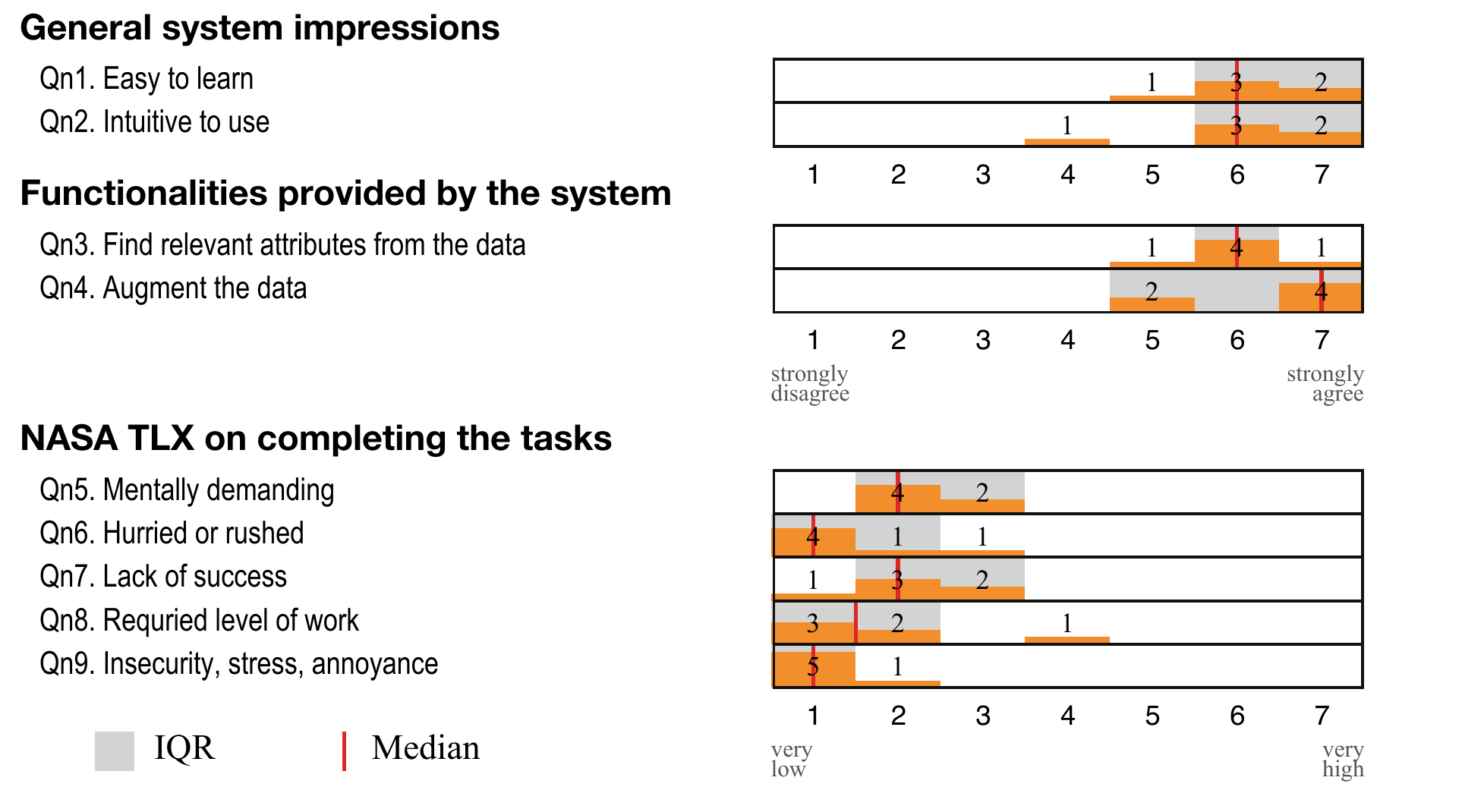}
    \vspace{-8mm}
    \caption{Participants' responses about the system (Qn1--Qn4, the higher the better) and the overall effort (Qn5--Qn9, the lower the better).}
        \vspace{-3mm}

    \label{fig:survey_ratings}
\end{figure}


\subsection{Result and Analysis}

\subsubsection{Task Performance}

To assess \textbf{H1}, we reviewed the number of attributes that participants correctly joined in Task 1.
Five out of six participants properly joined all four expected attributes.
Only one participants missed one attribute when joining the number of awards received by the director.
This indicates that in general participants were able to use \sys to accurately add new data attributes.

Next, to assess \textbf{H2}, we similarly reviewed the \textit{R Squared} score improvement that participants achieved by augmenting additional data in Task 2.
On average, participants improved the \textit{R Squared} Score by $0.048$ ($\sigma=0.014$), starting with a baseline \textit{R Squared} Score of $0.292$. 
The area and population of the county are two most frequently joined attributes across all the participants, which is reasonable as these two attributes are usually the most influential for predicting unemployment rate. 

As a result of our analysis, we \textbf{accept H1} because all participants were able to successfully use \sys to explore and identify the correct attributes for data joining in Task 1.
However, we \textbf{neither accept nor reject H2} as the study results did not contain clear evidence of users being able to improve the predictive quality of a regression model in Task 2 beyond trivial improvements.


\subsubsection{User satisfaction}

We use the post-study questionnaire to assess user satisfaction of the data augmentation process in \sys.
\autoref{fig:survey_ratings} shows the results of the questionnaire responses.
Although the participant size is too small to infer statistical significance using quantitative analysis, we do observe that participants found \sys easy and intuitive to use based on the likert scale user ratings (\autoref{fig:survey_ratings}, \textbf{Qn1,2}).
Participants were also satisfied with the two main functionalities provided by \sys, finding relevant attributes from the data and joining additional data (\autoref{fig:survey_ratings}, \textbf{Qn3,4}).
The mean ratings of \textbf{Qn1--4} were all 6 or above.
Furthermore, in analyzing user satisfaction related to the overall process (\autoref{fig:survey_ratings}, \textbf{Qn5--9}), we found that participants were generally satisfied with the process of conducting the tasks, as the mean ratings of \textbf{Qn5--9} are all 2 or lower.

\subsubsection{Qualitative Feedback}

To assess \textbf{H1} and \textbf{H2} from a qualitative perspective, we analysed participants’ descriptive feedback collected from post-study interviews. We also observed participants' workflow in using \sys from audio and video recordings of their computer.
We report the following three main qualitative user responses from the study:

\vspace{3pt} \noindent \textbf{Intuitive workflow:}
All the participants found \sys's workflow intuitive to find and augment relative data attributes.
The primary justification for \sys's intuitiveness is that \sys provides a visual interface that is easy to infer and for users to search and add relevant attributes from the data.
As \textit{P6} noted, ``\textit{I liked the interface of this tool because it was easy to navigate and add/remove columns to the base data}.''
The intuitiveness helped most participants to easily perform the requested tasks, especially Task 1.
For example, \textit{P1} described his experience in doing Task 1 as ``\textit{just looked at the question and implemented it}'' and ``\textit{it was pretty straight forward}.''

\vspace{3pt} \noindent \textbf{Two major strategies in model construction:}
When conducting Task 2, we observed two common strategies used by the participants: (1) searching for relevant attributes based on existing or prior knowledge and (2) exploring all possible combinations of available attributes to find one that results in improvement in model performance.
Some participants relied on their existing domain knowledge for augmenting the data with new attributes.
This was summarized nicely by \textit{P5}, ``\textit{I was actually using my domain knowledge. Thinking which factors can improve unemployment rate predictions, and then tried to select the variables based on the choices I had. This actually helped me improve the model's performance}.''
However, a few participants who may lack prior knowledge about the data, instead explored the dataset and tried out all possible combinations of attributes to find a model that is a significant improvement over the model trained on the base data.
``\textit{I selected columns one by one to see the change of R-squared score. By doing so, I was able to filter the variables that decrease R-square. In the end, I was able to get a relatively high R-squared model}'' (\textit{P6}).
The rest of the participants adopted a combination of both strategies.
For example, \textit{P2} described his strategies as follows, ``\textit{Thought a bit about what could boost performance, then I did some trial and error (manual feed-forward selection of attributes)}.''

\vspace{3pt} \noindent \textbf{Need for more powerful system features for expert users:}
Participants with visual analytics backgrounds expected editable visualizations to represent or encode attribute distribution differently.
``\textit{It would be cool to add additional ways of visualizing attributes}'' (\textit{P2}).
What's more, participants with database backgrounds wanted more details about queries of fetching data to increase flexibility.
``\textit{Maybe consider providing the real query of fetching the dataset to the expert users to give them more clear sense and allow them to change the query}” (\textit{P1}).
These suggestions are valuable for guiding our further improvement in generalizing \sys to tackle various problem domains.



\subsection{Limitations}

While our user study results \textbf{support H1} but \textbf{neither accept nor reject H2}, the results should be read in the light of the study limitations.
The number of the participants for our study are limited, which might be a confounding point of the results.
As we conducted the study online, the uncertainty existing in the online setup, such as internet connection, might also confound the results.
To improve the performance of the system and the fluidity of user interaction, we limited the number of related attributes that participants could fetch for each parent attribute.
Although it did help participants with a smoother user experience during the study, participants also complained about the restriction as it limited their performance of tasks, especially Task 2.
Despite the limitations, the user study does help us to better identify the limitations in our system design and improve the system with new features.



\section{Discussion}
\label{sec:discussion}

Through our user study and usage scenarios, we have demonstrated the efficacy of information foraging using knowledge graphs within visual analytics systems.  The generality of \sys suggests that data augmentation could be added into traditional visual analytics system workflows to improve the outcome of any embedded task.  Many fruitful avenues of research arise when considering how to apply information foraging to the full space of usage scenarios which are served by visual analytics systems.

\subsection{Mental Models of Data Augmentation}

Knowledge graphs can be difficult to reason about, especially when they are used to find data corresponding to an entire column of a dataset.  In \sys, we address this by showing the user previews of the joined dataset, as well as visualizations of the distribution of joined data and an example of the portion of the knowledge graph that is used to construct a single value (see Figure~\ref{fig:through_join}).  This approach was based on our conversations with designers of visual analytics applications for building predictive models.  A better understanding of the user's mental model of the knowledge graph and the joining process is needed to generalize this approach to other usage scenarios.  We generally aimed to hide the complexity of the underlying knowledge graph, but it may be the case that more direct exploration of the knowledge graph is helpful for some cases.

\subsection{Sparsity in Knowledge Graphs}

\label{sec:sparsity}

In this work, we rely on the high quality of Wikidata to find attributes that have sufficient support for the user's dataset.  But in many cases, only 10-15\% of the rows of a dataset will have joinable data for a given attribute.  Because knowledge graphs do not hold schematic data, they can suffer from data sparsity issues.  We feel that the experiments in this work show that there are many popular data types, including geographic entities like countries or states, that have sufficient robust data to discover attributes with full support on a user's dataset.  In situations where the attribute of interest is only present for a small percent of rows of the dataset, we suggest two possible solutions.  First, the additional data can still be useful as additional information injected into the dataset, and using imputation or setting reasonable default values (using a tool such as the \textit{data tamer}~\cite{Stonebraker2013} can make that information usable by downstream analyses.  Second, a user can explore the reduced dataset in which the sparse attribute is present, and if it results in promising analysis, they can use \sys to search for other more complete attributes that might be correlated with that sparse attribute, based on their domain expertise.  Lastly, we would like to point out that knowledge graphs are not read-only structures, and \sys may help point out where it would be valuable for an organization to invest in recording more data.

\subsection{Design Space for Interactions with Knowledge Graphs}

In \sys, the channel between the user and the knowledge graph is limited to the two subroutines shown in section~\ref{sec:technique}, which allow the user to see lists of related attributes and then construct queries given that information.  However, there is more potential to improve the user's control over the process and address edge cases by expanding the set of interactions between user and knowledge graph.  Automated processes in \sys could be replaced by collaborations between user and system.  

For example, ambiguities in the entity resolution used to retrieve a list of related attributes can be solved by user interaction.  A list of countries could refer to the governmental entities they describe, or they could refer to the national soccer teams participating in the World Cup; there will always exist cases where disambiguating the type of an attribute will require domain expertise from a user.  

The user could also benefit from more fine-grained control over the process of building aggregation queries.  Users may want to join timestamped data, which would necessitate some specification from the user of how to parse and interpret temporal columns.  Spatial data offers an additional potential, as the user might want to use geographical data in their dataset to search for the closest weather station or other geographically tagged entity.  There are many types of queries that might necessitate different user interactions than have been used in previous visual analytics systems.

\section{Conclusion}
\label{sec:conclusion}
In this work, we presented \sys, a visual analytics system for explorative information foraging using knowledge graphs.  Knowledge graphs offer a wealth of information on a broad array of topics that could be used to improve the outcome of embedded tasks of visual analytics systems.  In our usage scenarios, we demonstrated that the data gathered through \sys could result in better predictive models and better insight generation on two datasets.  And in our experiment, we showed that \sys is simple to learn and use, as all six of our participants were able to effectively explore and join data on multiple datasets within a 60 minute session.

\acknowledgments{
We thank our collaborators in DARPA's Data Driven Discovery of Models (d3m) program, as well as the reviewers for their helpful feedback.  This work was supported by National Science Foundation grants IIS-1452977, OAC-1940175, OAC-1939945, DGE-1855886, and 1841349, as well as DARPA grant FA8750-17-2-0107.
}
\bibliographystyle{abbrv-doi}

\bibliography{main}
\end{document}